%% file: nlo-journalversion.tex
\documentclass[prd,aps,preprint,tightenlines,superscriptaddress,nofootinbib,showpacs,showkeys,floatfix]{revtex4}
\usepackage{amsfonts}
\usepackage{array}
\usepackage{epsfig}
\usepackage{rotating}
\usepackage{multirow}

\usepackage{amsmath}    
\usepackage{verbatim}   
\usepackage{color}      
\usepackage{colordvi}

\usepackage{subfigure}  
\usepackage{hyperref}   

\usepackage{pstricks,rotating, graphicx}
\usepackage{cancel}

\usepackage{tablefootnote}  
\usepackage[normalem]{ulem}

\newcommand{\GeV}{\ensuremath{\,\mathrm{GeV}}}

\newcommand{\msbar}{$\overline{\mathrm{MS}}\, $}

\graphicspath{{./Pics/}}
\DeclareGraphicsExtensions{.pdf}

\usepackage{pslatex}
\usepackage{txfonts}
\usepackage[latin1]{inputenc}
\usepackage[T1]{fontenc}

\begin{document}

\begin{titlepage}
\thispagestyle{empty}
\noindent
DESY 18-026 \\
DO-TH 17/28 \\
\hfill
February 2018 \\
\vspace{1.0cm}

\begin{center}
  {\bf \Large
    NLO PDFs from the ABMP16 fit
  }
  \vspace{1.25cm}

 {\large
   S.~Alekhin$^{\, a,b}$,
   J.~Bl\"umlein$^{\, c}$,
   and
   S.~Moch$^{\, a}$
   \\
 }
 \vspace{1.25cm}
 {\it
   $^a$ II. Institut f\"ur Theoretische Physik, Universit\"at Hamburg \\
   Luruper Chaussee 149, D--22761 Hamburg, Germany \\
   \vspace{0.2cm}
   $^b$Institute for High Energy Physics \\
   142281 Protvino, Moscow region, Russia\\
   \vspace{0.2cm}
   $^c$Deutsches Elektronensynchrotron DESY \\
   Platanenallee 6, D--15738 Zeuthen, Germany \\
 }
  \vspace{1.4cm}
  \large {\bf Abstract}
  \vspace{-0.2cm}
\end{center}
We perform a global fit of parton distribution functions (PDFs) together with 
the strong coupling constant $\alpha_s$ and the quark masses $m_c$, $m_b$ and $m_t$ 
at next-to-leading order (NLO) in QCD.
The analysis applies the \msbar\ renormalization scheme for $\alpha_s$ and all quark masses. 
It is performed in the fixed-flavor number scheme for $n_f=3, 4, 5$ and 
uses the same data as the previous fit of the ABMP16 PDF at next-to-next-to-leading order (NNLO).
The new NLO PDFs complement the set of ABMP16 PDFs and are to be used consistently with NLO QCD predictions for hard
scattering processes.
At NLO we obtain the value $\alpha_s^{(n_f=5)}(M_Z) = 0.1191 \pm 0.0011$ 
compared to $\alpha_s^{(n_f=5)}(M_Z) = 0.1147 \pm 0.0008$ at NNLO.

\end{titlepage}

Parton distribution functions (PDFs) are an indispensable ingredient in 
theory predictions for hadronic scattering processes within perturbative QCD.
Currently, the state-of-art calculations for many standard-candle processes 
at the Large Hadron Collider (LHC) and elsewhere are based 
on the QCD corrections up to the next-to-next-to-leading order (NNLO) 
in the strong coupling constant $\alpha_s$~\cite{Accardi:2016ndt}.
In order to match this theoretical accuracy the PDFs and other input
parameters such as $\alpha_s$ and the quark masses $m_c$, $m_b$ and $m_t$ 
also have to be determined at the same order of perturbation theory, that
is with account of the NNLO QCD corrections.
In many instances, however, the Wilson coefficient functions or 
hard partonic scattering cross sections are known to the 
next-to-leading order (NLO) only.
This concerns in particular Monte-Carlo studies at the LHC. 
Then, to meet the consistency requirements, NLO PDFs and the
respective NLO values for $\alpha_s$ and the heavy-quark masses are to be used. 
The NLO fit of PDFs is therefore of immediate practical use and also provides a very good
consistency check of the perturbative stability of QCD calculations. 

In this article we describe the NLO version of the recent ABMP16 PDF fit, 
i.e., the NLO analysis, which applies the \msbar\ scheme for $\alpha_s$ and all heavy-quark masses. 
It uses the same data, their uncertainty treatment and the general theoretical
framework, e.g., the fixed-flavor number scheme for $n_f=3, 4$ and $5$, 
as in the previous fit of the ABMP16 PDF at NNLO.
The only difference resides in the order of the perturbative corrections 
to the QCD evolution equations and for the Wilson coefficients, 
which are now limited to NLO accuracy. 
Due to the obvious correlations of the various parameters in 
the PDFs with the value of $\alpha_s$ and those of the quark masses $m_c$, $m_b$ and $m_t$, 
all quantities are extracted simultaneously from the global fit 
following our previous analyses~\cite{Alekhin:2012ig,Alekhin:2013nda,Alekhin:2017kpj}.
The article discusses in detail the differences in their determinations at NLO and NNLO accuracy.
Specific attention is paid to the treatment of power corrections in the
description of deep-inelastic scattering (DIS) data, i.e. higher-twist
effects which are relevant beyond the leading twist collinear factorization approximation.
The final fit results are made available as data grids for use with the {\tt LHAPDF} library (version 6)~\cite{Buckley:2014ana} 
and the features of the various grids are briefly discussed.

\bigskip

\input{table-chi2.tex}

The values of $\chi^2$ obtained in the present analysis for various data sets 
are listed in Tab.~\ref{tab:chi2} in comparison with the earlier ones for the NNLO ABMB16 fit. 
The overall quality of the data description does not change dramatically between the NLO
and the NNLO versions, where the former features a somewhat bigger total value
of $\chi^2$.
Of course, the theoretical description at NNLO accuracy comes with a
significantly reduced theoretical uncertainty due to variations 
of the factorization and renormalization scales compared to the NLO one.
Nevertheless, for specific scattering reactions the NNLO corrections are crucial 
for the respective data sets.
This holds in particular for the $c$-quark and, to a lesser extent, 
for $b$-quark production in DIS and for hadronic $t$-quark pair-production, 
which constrain the heavy-quark masses and which are 
fitted together with $\alpha_s$ simultaneously with the PDFs.
The theoretical description at NNLO accuracy is also essential for the 
parameters of the higher (dynamical) twist, which contribute additively to the leading twist.
The $x$-dependent twist-four contributions to the longitudinal and transverse DIS cross
sections have been determined in the NNLO version (cf. Tab.~VIII in Ref.~\cite{Alekhin:2017kpj})
and their central values are kept fixed in the present analysis.
Also other fitted parameters like the data set normalizations are taken over unchanged 
from the NNLO analysis (cf. Tab.~I in Ref.~\cite{Alekhin:2017kpj}).
This provides a better consistency between the PDF sets obtained with different
theoretical accuracy. 
At the same time the uncertainties in the normalization and higher twist parameters 
are computed in the same way as in the NNLO fit~\cite{Alekhin:2017kpj}, by propagation of the ones in experimental data 
and simultaneosly with other fit parameters in order to take into account their correlations and, therefore,
provide a consistent uncertainty treatment in the NLO and NNLO fits.
Therefore, the uncertainties obtained for the data normalization and the twist-four contributions at NLO are only 
marginally different from those reported at NNLO in Tabs.~I and VIII in Ref.~\cite{Alekhin:2017kpj}.

Closer inspection of the $\chi^2$-values in Tab.~\ref{tab:chi2} reveals 
the largest differences between NLO and NNLO for the fixed-target DIS data, 
which can be explained by the kinematic coverage of this data sample, 
which is predominantly in the low-$Q^2$ region.
For the Drell-Yan (DY) data the impact of the NNLO QCD corrections is less
pronounced and the corresponding improvement in the value of $\chi^2$ is small.
For the heavy-quark production data on the other hand the trend is not
uniform, i.e., for some data sets the $\chi^2$-values at NLO are larger and vice versa for others. 
In this context it is worth noting, that 
to a certain extent the impact of missing NNLO terms in the NLO fit 
is compensated by tuning the values of heavy-quark masses.

\begin{figure}[t!]
    \centerline{
    \includegraphics[width=16.0cm]{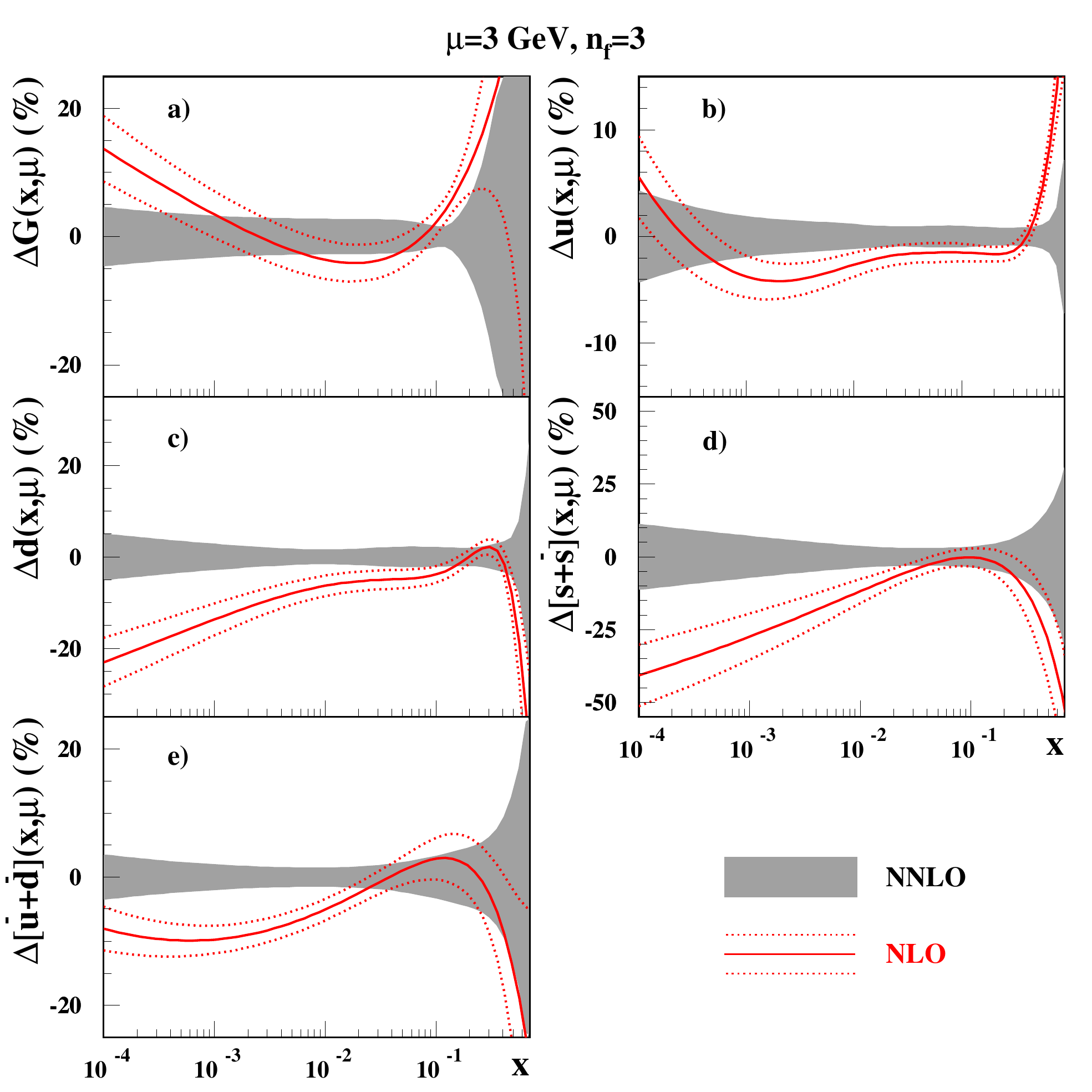}
}
\vspace*{-2mm}
\caption{\small
  \label{fig:pdfs}
  The $1\sigma$ band for the $n_f=3$ flavor NNLO ABM16 PDFs~\cite{Alekhin:2017kpj} 
  for {\bf a}) gluon, {\bf b}) up-quarks 
  {\bf c}) down-quarks 
  {\bf d}) the symmetrized strange sea 
  and {\bf e}) the non-strange sea,
  at the scale of $\mu = 3$~GeV versus $x$ (shaded area) compared with the relative difference of those PDFs 
  to the NLO ABMP16 ones obtained in the present analysis (solid lines). 
  The dotted lines display $1\sigma$ band for the NLO PDFs. 
}
\end{figure}

The ABMP16 PDF sets at NLO and NNLO are compared in Fig.~\ref{fig:pdfs} for the
case of $n_f=3$ flavors at the scale $\mu=3$~GeV. 
Both sets are based on the same flexible parametrization used in Ref.~\cite{Alekhin:2017kpj}.
For the gluon PDF, we see in Fig.~\ref{fig:pdfs}a 
that the NLO PDFs are larger by about $15\%$ in the small-$x$ and the large-$x$ region,
i.e., for $x\lesssim 10^{-4}$ and $x\gtrsim 0.3$, respectively.
In these kinematic regions for example the DIS coefficient functions 
receive systematically large corrections at higher orders, 
which need to be compensated by the gluon PDF 
if the fit is performed at NLO accuracy.
The $u$-quark PDF in Fig.~\ref{fig:pdfs}b does not show any big
changes, except for large $x\gtrsim 0.6$, 
while the $d$-quark PDF Fig.~\ref{fig:pdfs}c at NLO is smaller in the
entire range $x\lesssim 10^{-1}$ and decreasing more than $20\%$ for 
$x\lesssim 10^{-4}$.
A similar observation holds for the strange sea displayed in
Fig.~\ref{fig:pdfs}d, which is smaller by even $50\%$ for 
$x\lesssim 10^{-4}$ at NLO, however, the PDF uncertainties for this quantity are
correspondingly larger.
On the other hand, the non-strange sea in Fig.~\ref{fig:pdfs}e 
does not show big relative differences between NLO and NNLO. There is only a
slight decrease of the NLO result by $5\%$ to $10\%$ for 
$x\lesssim 10^{-2}$.
The small-$x$ sea iso-spin asymmetry $\bar d - \bar u$ at NLO goes lower 
than the NNLO one, as can be seen from a comparison of Figs.~\ref{fig:pdfs}b) 
and \ref{fig:pdfs}c).
This reflects the impact of the NNLO corrections on the data 
for Drell-Yan production, which drive this asymmetry in our fit.

\bigskip

\input{table-cut.tex}

\begin{figure}[h!]
    \centerline{
    \includegraphics[width=10.5cm]{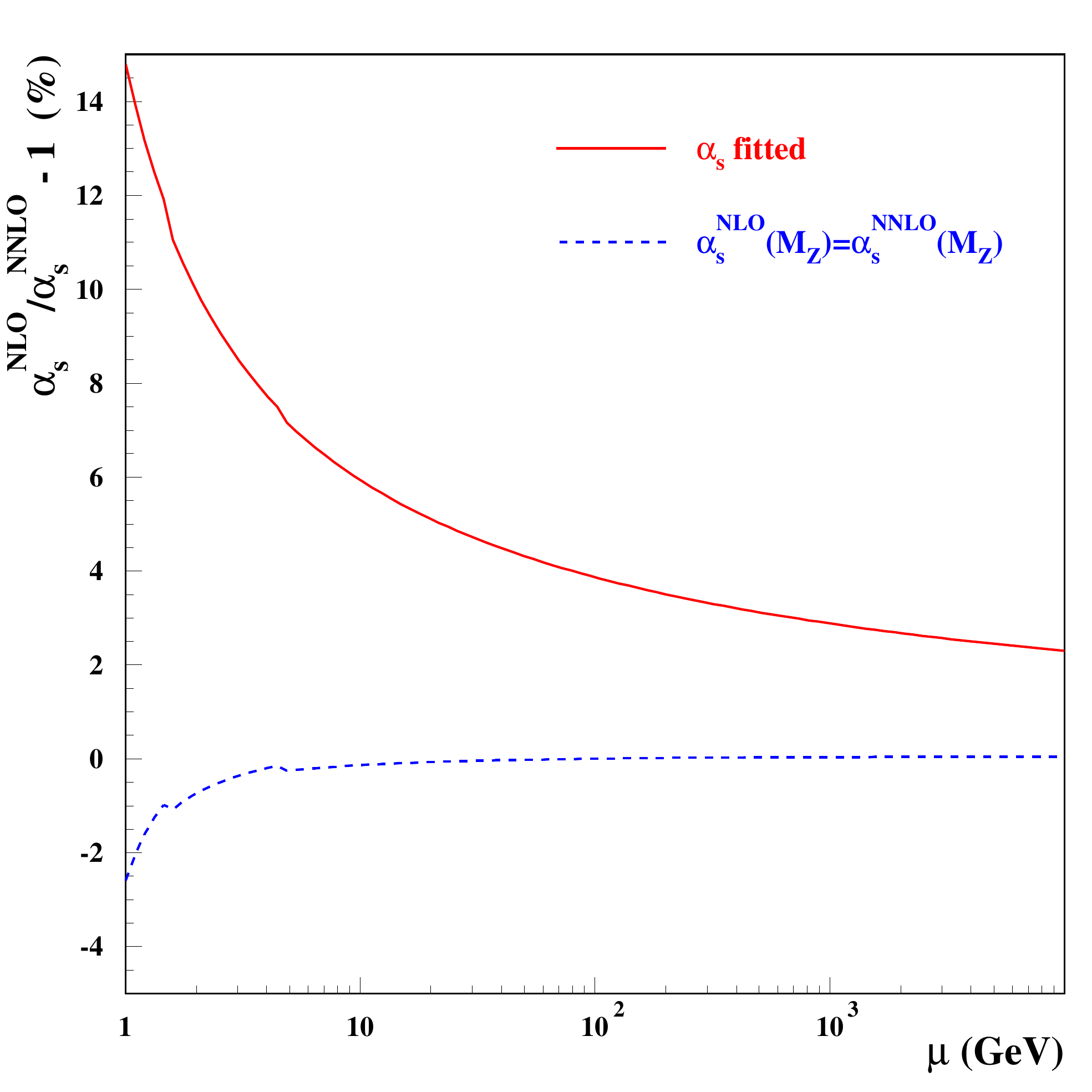}
}
\vspace*{-2mm}
  \caption{\small
    \label{fig:alps}
    The relative difference between $\alpha_s^{\rm NLO}(\mu)$ and $\alpha_s^{\rm NNLO}(\mu)$ as 
    a function of the renormalization scale $\mu$.
    The solid line denotes the results of the present analysis (NLO) and of
    Ref.~\cite{Alekhin:2017kpj} (NNLO), 
    the dashed line displays the ones derived using matching 
    $\alpha_s^{\rm NLO}=\alpha_s^{\rm NNLO}$ at the scale $M_Z$ .
  }
\end{figure}

The value of $\alpha_s(M_Z)$ at the scale of $Z$-boson mass $M_Z$
obtained in the present analysis at NLO is larger than the 
one obtained in the NNLO variant of ABMP16 fit~\cite{Alekhin:2017kpj}, 
the relative difference amounting to about $4\%$, 
which is well comparable to the estimated margin due to variations 
of the factorization and renormalization scales, cf. Ref.~\cite{Blumlein:1996gv}.
In the scheme with $n_f=5$ light flavors we find 
\begin{eqnarray}
  \label{eq:as}
  \alpha_s^{{\rm NLO}}(M_Z)&=&  0.1191 \pm 0.0011
  \, ,
  \\
  \nonumber
  \alpha_s^{{\rm NNLO}}(M_Z)&=&  0.1147 \pm 0.0008
  \, ,
\end{eqnarray}
as listed in the first line of Tab.~\ref{tab:cuts} together with the kinematic 
cuts imposed on the DIS data.
The description of DIS data at those low values of $Q^2$ and $W$ for the invariant mass of the hadronic system, 
where $W^2 = M_P^2 + Q^2 (1-x)/x$ with the proton mass $M_P$, requires modeling of the higher-twist terms.
This has been discussed extensively in the ABMP16 analyses at NNLO~\cite{Alekhin:2017kpj}. 
Following the theoretical framework there, the fitted twist-four 
contributions to the longitudinal and transverse DIS cross sections
have been used to determine the value of $\alpha_s^{{\rm NLO}}(M_Z)$ in Eq.~(\ref{eq:as}).
Alternatively, one can impose cuts both on $Q^2$ and $W^2$ to eliminate data
from the kinematic regions most sensitive to the higher-twist terms. 
Then, the fit can be performed with all higher-twist terms set to zero and the
results are shown in Tab.~\ref{tab:cuts}.
These variants of the fit with substantially higher cuts on $Q^2$ and $W^2$
and higher-twist terms set to zero display very good stability of the value
of $\alpha_s(M_Z)$, both at NLO and NNLO, and therefore very good consistency
of the chosen approach. 

Finally, in one of the variants of present analysis we impose the low cuts on $Q^2$ and $W^2$ 
from the first line of Tab.~\ref{tab:cuts}, while fitting also the twist-four contributions.
This gives an improvement in the value of $\chi^2$ equal to 86 and 
$\alpha_s^{{\rm NLO}}(M_Z) = 0.1227 \pm 0.0011$, 
which is a slightly larger value than those quoted in Tab.~\ref{tab:cuts} 
for the fits with higher cuts on $Q^2$ and $W^2$. 
The magnitude of these shifts in $\alpha_s(M_Z)$ may also be considered as an
indication for the limitations of the NLO approximation.

Nevertheless, the observed difference between $\alpha_s^{{\rm NLO}}(\mu)$ and $\alpha_s^{{\rm NNLO}}(\mu)$ is quite essential, 
particularly at small scales $\mu$, where the NLO and NNLO results differ by more than $10\%$, 
as illustrated in Fig.~\ref{fig:alps} for a wide range of scales. 
This difference is to a great extent responsible for the perturbative stability
of QCD calculations at the hard scales currently probed in scattering processes at colliders.
Asymptotic freedom in QCD, i.e. stability of theoretical predictions under
higher order perturbative corrections requires very large scales. 
On the other hand, for realistic kinematics including experiments at the LHC 
a consistent setting of $\alpha_s(M_Z)$ is very important to achieve sensible
theoretical predictions.\footnote{
Recent reviews on determinations of $\alpha_s(M_Z)$, particularly those involving PDF fits, 
can be found in \cite{Alekhin:2016evh}, 
in Sec.~4 of Ref.~\cite{Accardi:2016ndt} and in Sec.~III.D of Ref.~\cite{Alekhin:2017kpj}.
Determinations of $\alpha_s(M_Z)$ in DIS and including jet cross section
measurements have been discussed in Ref.~\cite{Andreev:2017vxu}.}

In this context it is worth to mention the conventional choice  
$\alpha_s^{\rm NLO}(M_Z)=\alpha_s^{\rm NNLO}(M_Z)$, which is 
adopted as a part of PDF4LHC recommendations~\cite{Rojo:2015acz}
and employed in the CT14~\cite{Dulat:2015mca} and NNPDF~\cite{Ball:2017nwa} PDF fits. 
Under this assumption the value of $\alpha_s$ obtained at NLO 
is very close to the NNLO one in a wide range of scales, as shown in Fig.~\ref{fig:alps}.
As a result, such an approach has significant limitations when studying the convergence of the perturbative expansion, 
since the NLO predictions obtained with these PDF sets might be very similar to the NNLO ones simply due to the convention used.

\bigskip

The values for the heavy-quark masses obtained in the NLO and NNLO variants of
the ABMP16 analysis are given in Tab.~\ref{tab:hqmass}. 
The comparison of the difference between these values indicates again the
limitations of the NLO approximation. 
The shifts are more essential for $m_c$ and $m_t$, since 
the NNLO corrections are absolutely essential in order to achieve a good
description of the data on DIS $c$-quark production and $t$-quark hadro-production.
On the other hand, the data on DIS $b$-quark production are less precise, therefore the value of $m_b$ extracted from the fit suffers from  
the larger uncertainties  and is less sensitive to the impact of the NNLO corrections, cf. Tab.~\ref{tab:hqmass}.
\input{table-mhq.tex}

\begin{figure}[th!]
\centerline{
  \includegraphics[width=9.25cm]{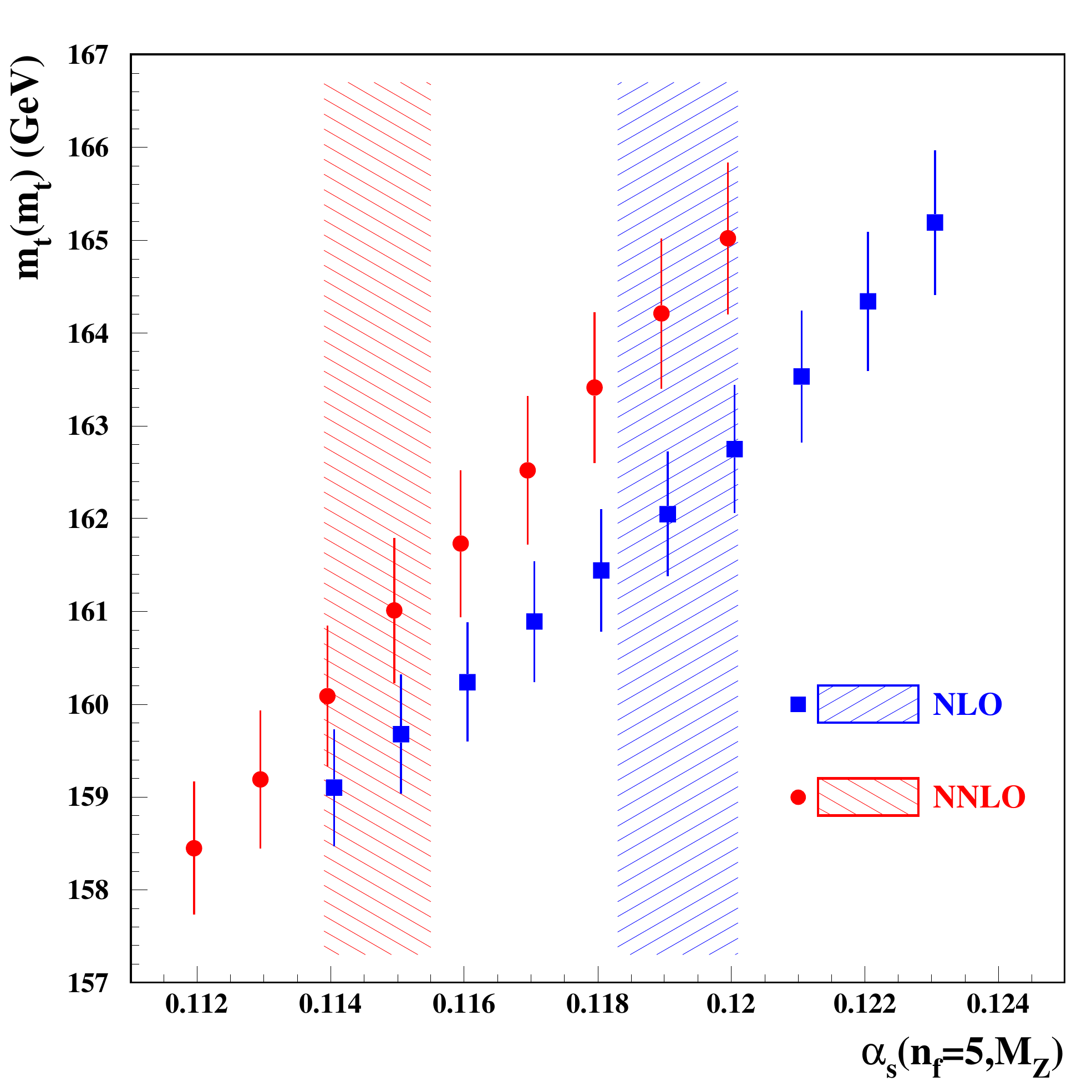}}
\caption{\small
  \label{fig:scanmt}
  The \msbar\ values of the $t$-quark mass $m_t(m_t)$ obtained in the
  variants of the NLO ABMP16 fit 
  with $\alpha_s^{(n_f=5)}(M_Z)$ fixed (squares) 
  in comparison to ones at NNLO (circles).   
  The left-tilted and right-tilted hatch represent the
  $1\sigma$ bands for $\alpha_s^{(n_f=5)}(M_Z)$ 
  obtained in the ABMP16 nominal fits at NLO and NNLO, respectively. 
  The points are slightly shifted left and right to prevent overlapping. 
}
\end{figure}

\begin{figure}[th!]
  \centerline{
    \includegraphics[width=9.25cm]{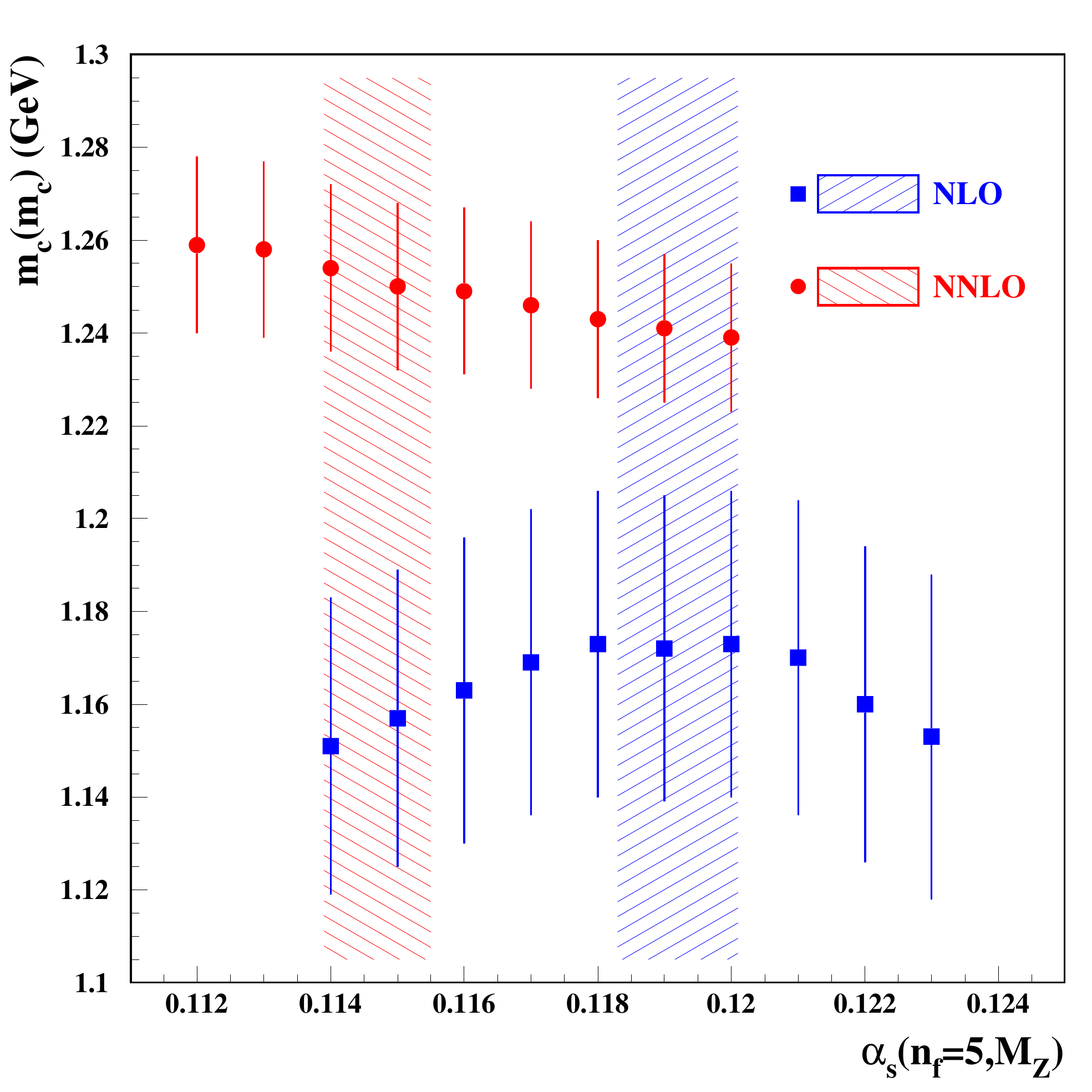}}
  \caption{\small
    \label{fig:scanmc}
    The same as in Fig.~\ref{fig:scanmt} for 
    the \msbar\ value of the $c$-quark mass $m_c(m_c)$.
  }
\end{figure}

In addition, the value of $m_t$ demonstrates strong correlation with the value of $\alpha_s$, since the 
Born cross section for $t{\bar t}$-production is proportional to $\alpha_s^2$, 
so that changes in the value of $\alpha_s$ induce shifts in 
fitted value of $m_t$~\cite{Alekhin:2013nda,Alekhin:2017kpj}.
This is quantified in Fig.~\ref{fig:scanmt}, 
where the values of $m_t$ determined in variants of the present analysis 
with fixed values of $\alpha_s$ demonstrate a nearly linear dependence on $\alpha_s$. 
It is interesting to note, though, that at NLO this dependence is somewhat 
shallower than for the similar fit at NNLO since due to important missing QCD corrections of ${\cal O}(\alpha_s^4)$ in the 
hadronic $t\bar{t}$-production cross section at NLO,  
$\sigma(t{\bar t})$ is less sensitive to the $\alpha_s$ variations at this order. 
For the same reason the NLO value of $m_t$ is substantially larger than the NNLO one. 
This comparison demonstrates the necessity for a consistent treatment of the
higher order corrections to the hard partonic scattering together 
with the parameter choices for $\alpha_s$ and $m_t$, 
especially in analysis of the $t\bar{t}$-production data. 
To that end, the PDFs from the variants entering Fig.~\ref{fig:scanmt} 
with the a fixed value of $\alpha_s(M_Z)$ in the range of $0.114 \div 0.123$ 
are made available as well.
It should be stressed, though, that those PDFs at preselected values of $\alpha_s(M_Z)$ 
are not providing the $\chi^2$ minimum in analysis of all the data. 

In contrast, the determination of the charm-quark mass $m_c$ is not very
sensitive to the value of $\alpha_s$ because the variation of $\alpha_s$ is
compensated by a change in the gluon distribution at small $x$~\cite{{Prytz:1993vr}}  
(cf. also Tab.~B in Ref.~\cite{Alekhin:2017kpj} for correlations between $\alpha_S$ and the quark masses).
Indeed, the fitted value of $m_c$ only changes by $\pm 20~{\rm MeV}$ 
for a variation of $\alpha_s$ in a wide range, cf. Fig.~\ref{fig:scanmc}. 
However, the NNLO corrections to the Wilson coefficents for DIS heavy-quark production still have significant impact on $m_c$, 
moving it up by $\sim 100~{\rm MeV}$ and reducing its uncertainty. 
The relatively weak correlation of $m_c$ with $\alpha_s$ in Fig.~\ref{fig:scanmc} 
is in contrast to the observed behavior in the MMHT14 PDF set~\cite{Harland-Lang:2015qea}, where the 
particular variable flavor number scheme applied causes a linear relation between 
$m_c$ and $\alpha_s$ in those fits 
and the charm-quark mass has been treated as a variable parameter and 
the resulting values of $\chi^2$ when fitting data on DIS $c$-quark production have been quantified.
See Sec.~3.2 and Tabs.~4 and 5 in Ref.~\cite{Accardi:2016ndt} for a review of the 
theoretical treatments of DIS $c$-quark production used in PDF fits
and Tab.~2 in Ref.~\cite{Harland-Lang:2015qea} as well as 
Tab.~12 in Ref.~\cite{Accardi:2016ndt} for the respective 
values of $m_c$, $\alpha_s$ and $\chi^2$ in the MMHT14 PDF set.

\bigskip

\input{table-asmz-mq-parameters-nlo}

The grids for the NLO PDFs obtained in the present analysis 
are accessible with the {\tt LHAPDF} library (version 6)~\cite{Buckley:2014ana}
and available for download under {\tt http://projects.hepforge.org/lhapdf}.
For a fixed number of flavors, $n_f=3, 4$ and $5$, we provide 
\begin{verbatim}
      ABMP16_3_nlo (0+29),
      ABMP16_4_nlo (0+29),
      ABMP16_5_nlo (0+29),
\end{verbatim}
which consist of the central fit (set 0) 
and additional 29 sets for the combined symmetric uncertainties in all 
parameters (PDFs, $\alpha_s$, $m_c$, $m_b$ and $m_t$). 
In each PDF set, the strong coupling $\alpha_s$ is taken in the corresponding
scheme, i.e., $\alpha_s^{(n_f=3)}$, $\alpha_s^{(n_f=4)}$ and $\alpha_s^{(n_f=5)}$ 
which can be related by the standard decoupling relations in QCD.
As usual, the PDF set with three light-quarks $n_f=3$, {\tt ABMP16\_3\_nlo}, 
is valid at all perturbative scales $\mu^2 \gtrsim 1$~GeV$^2$, 
while those with $n_f=4$ and $n_f=5$, {\tt ABMP16\_4\_nlo} and {\tt ABMP16\_5\_nlo}, 
are subject to minimal cuts in $\mu^2 \ge 3$~GeV$^2$ 
and $\mu^2 \ge 20$~GeV$^2$, respectively, cf.
 Ref.~\cite{Alekhin:2017kpj} for additional discussions.

For studies of LHC observables and their dependence on $\alpha_s$ 
we also provide NLO PDF grids for $n_f=5$ flavors 
with the central value of $\alpha_s^{(n_f=5)}(M_Z)$ fixed. 
The 10 sets cover the range $\alpha_s^{(n_f=5)}(M_Z)=0.114 
\div 0.123$ with a spacing of 0.001 and are denoted as 
\begin{verbatim}
      ABMP16als114_5_nlo (0+29),
      ABMP16als115_5_nlo (0+29),
      ABMP16als116_5_nlo (0+29),
      ABMP16als117_5_nlo (0+29),
      ABMP16als118_5_nlo (0+29),
      ABMP16als119_5_nlo (0+29),
      ABMP16als120_5_nlo (0+29),
      ABMP16als121_5_nlo (0+29),
      ABMP16als122_5_nlo (0+29),
      ABMP16als123_5_nlo (0+29),
\end{verbatim}
where the value of $\alpha_s^{(n_f=5)}(M_Z)$ has been fixed as indicated in the file names. 
These grids are determined by re-fitting all PDF parameters for the individual choices of $\alpha_s$, 
which for technical consistency remains a formal parameter in the fit, but with greatly suppressed uncertainty.

As the heavy-quark masses $m_c(m_c)$, $m_b(m_b)$ and $m_t(m_t)$ have been
fitted their numerical values vary for each of the 29 PDF sets and 
cross section computations involving heavy quarks have to account for this.
For reference we list in Tab.~\ref{tab:parameters} the heavy-quark masses 
in the ABMP16 grids and the values of $\alpha_s^{(n_f=5)}(M_Z)$.
These values can also be easily retrieved within the {\tt LHAPDF} library framework.  
The bottom- and the top-quark pole masses, $m_b^{\rm pole}$ and $m_t^{\rm pole}$,
which are required for the on-shell scheme are also provided in Tab.~\ref{tab:parameters}.
In particular, for computations with the central ABMP16 set at NLO the values 
$m_b^{\rm pole} = 4.488~\GeV$ and $m_t^{\rm pole} = 171.44~\GeV$ should be used. 
\footnote{In the matching of $m_c(m_c)$ to the on-shell scheme 
$m_c^{\rm pole}$ acquires large QCD corrections up to N$^3$LO~\cite{Marquard:2015qpa},  
therefore use of $m_c^{\rm pole}$ is problematic in this context~\cite{Accardi:2016ndt}.}

Finally, we also provide the results of the variants with no constraints 
on the fit parameters, in particular on the higher-twist terms, which are 
extracted at NLO as well. 
For $n_f=3, 4$ and $5$ flavors these are the sets 
\begin{verbatim}
      ABMP16free_3_nlo (0+29),
      ABMP16free_4_nlo (0+29),
      ABMP16free_5_nlo (0+29),
\end{verbatim}
which come, as discussed above, essentially with larger values for $\alpha_s(M_Z)$ and $m_t(m_t)$,  
e.g., $\alpha_s^{(n_f=5)}(M_Z)=0.1227$ and $m_t(m_t)=164.47~\GeV$. 

\bigskip

\input{table-crs}

The benchmark cross sections for the Higgs-boson and top-quark pair production at the LHC at NLO
and NNLO with consistent use of the PDF sets obtained in the 
present analysis are given in Tab.~\ref{tab:crs}.
The quoted errors denote the PDF and $\alpha_s$ uncertainties 
derived from the uncertainties in the experimental data. 
Thus, they are of similar size at NLO and NNLO.

The Higgs boson cross section $\sigma(H)$ is computed in the effective theory
in the limit $m_t \to \infty$, but with full $m_t$ dependence in the
Born cross section, based on the NNLO results of Refs.~\cite{Harlander:2002wh,Anastasiou:2002yz,Ravindran:2003um}. 
The NLO value of $\sigma(H)$ is about 20\% smaller than the NNLO one due to
missing large perturbative corrections, which are only partially compensated by a larger value of $\alpha_s$.
In the effective theory the Born cross section for $\sigma(H)$ 
is proportional to $\alpha_s^2$, so that the variant of the NLO fit with the
larger value for the strong coupling, $\alpha_s^{{\rm NLO}}(M_Z) = 0.1227$, 
gives a NLO cross section increased by 5\%, i.e. $\sigma(H)=35.2 \pm 0.58$~pb with
the PDF set {\tt ABMP16free\_5\_nlo} compared to 
$\sigma(H)= 33.59 \pm 0.58$~pb with the PDF set {\tt ABMP16\_5\_nlo} in Tab.~\ref{tab:crs}. 

The inclusive cross section $\sigma(t{\bar t})$ for top-quark pair production
uses Ref.~\cite{Aliev:2010zk} based on Refs.~\cite{Baernreuther:2012ws,Czakon:2012zr,Czakon:2012pz,Czakon:2013goa}.
In this case, the NLO and NNLO values of $\sigma(t{\bar t})$ for the range of
center-of-mass energies explored at the LHC are similar, since those data have been included in both
fits and are accommodated by the corresponding changes in the value of $\alpha_s$
and the top-quark mass $m_t$, cf. Fig.~\ref{fig:scanmt} and Tab.~\ref{tab:hqmass}.

\bigskip

In summary, we have completed the determination of the ABMP16 PDF sets 
at those orders of perturbation theory, which are currently of
phenomenological relevance, i.e., at NLO and NNLO.
Essential input in the ABMP16 analysis has been the final HERA DIS combination data from run I+II, 
which has consolidated the available world DIS data.
In addition, several new data sets from the fixed-target DIS together   
with recent LHC and Tevatron data for the DY process and for the top-quark hadro-production
have been used.

We have discussed the features of the NLO extraction of PDFs, which in
general, have a few limitations due to lacking constraints of the higher order
Wilson coefficients and we have emphasized the consistent use of PDFs 
and an order-dependent value of $\alpha_s(M_Z)$, which is absolutely crucial
because of correlations. 
The same holds, to a lesser extent, in collider processes 
also for the values of the heavy-quark masses used.

The ABMP16 PDFs establish the baseline for high precision analyses of LHC data
from run I and run II, and the NLO variant is now available for computing 
cross sections of scattering processes with multi-particle final states, 
for which the NNLO QCD corrections will not be available in the foreseeable
future, or for Monte Carlo studies.
Precision analyses of LHC data, however, will always require analyses to NNLO
accuracy in QCD. 
This will become even more important with the arrival of the data from the
high luminosity runs.

\bigskip

{\bf{Acknowledgments:}}\qquad 
We would like to thank K.~Lipka and O.~Zeniaev for cross-checks of the preliminary version of the {\tt LHAPDF} grids derived from this analysis.
This work has been supported by Bundesministerium f\"ur Bildung und Forschung (contract 05H15GUCC1).


\end{document}

%% file: table-chi2.tex
\begin{table}[t!]
\renewcommand{\arraystretch}{1.3}
\begin{center}                   
  \begin{tabular}{|l|c|c|c|c|}
    \hline
Experiment&Process & $NDP$&\multicolumn{2}{c|}{$\chi^2$}
\\
\hline
\multicolumn{1}{l}{} & \multicolumn{1}{c}{} & \multicolumn{1}{c}{} &\multicolumn{1}{|c|}{NLO} & \multicolumn{1}{c|}{NNLO}  \\
\cline{4-5} 
\multicolumn{1}{l}{\bf DIS} & \multicolumn{1}{c}{} & \multicolumn{1}{c}{} &\multicolumn{1}{c}{} & \multicolumn{1}{c}{}  \\ 
\hline
HERA~I+II &$e^{\pm}p \rightarrow e^{\pm} X$ & 1168 & 1528 & 1510 
\\
   &$e^{\pm}p \rightarrow \overset{(-)}{\nu} X$ &  &  &  
\\
\hline
Fixed-target (BCDMS, NMC, SLAC) &$l^\pm p \rightarrow l^\pm X$ & 1008 & 1176 & 1145
\\
\hline
\multicolumn{1}{l}{\bf DIS heavy-quark production} & \multicolumn{1}{l}{} & \multicolumn{1}{c}{} &\multicolumn{1}{c}{} & \multicolumn{1}{c}{}  \\ 
\hline
HERA~I+II &$e^{\pm}p \rightarrow e^{\pm}c X$ & 52 & 58 & 
66~\footnote{This value corrects a misprint in Table~V of Ref.~\cite{Alekhin:2017kpj}.}
\\
\hline
H1, ZEUS  &$e^{\pm}p \rightarrow e^{\pm}b X$ & 29 & 21 & 21 
\\
\hline
Fixed-target (CCFR, CHORUS, NOMAD, NuTeV) & $\overset{(-)}{\nu} N \rightarrow \mu^{\pm} c X$   &  232 & 173  & 178   
\\
\hline
\multicolumn{1}{l}{\bf DY} & \multicolumn{1}{c}{} & \multicolumn{1}{c}{} &\multicolumn{1}{c}{} & \multicolumn{1}{c}{}  \\ 
\hline
ATLAS, CMS, LHCb &$p p \rightarrow W^\pm X$ & 172 & 229 & 223
\\
 &$p p \rightarrow Z X$ &  & & 
\\
\hline
Fixed-target (FNAL-605, FNAL-866) &$p N \rightarrow \mu^+ \mu^- X$ & 158 & 219 & 218
\\
\hline
\multicolumn{1}{l}{\bf Top-quark production} & \multicolumn{1}{l}{} & \multicolumn{1}{c}{} &\multicolumn{1}{c}{} & \multicolumn{1}{c}{}  \\ 
\hline
ATLAS, CMS & $pp \rightarrow tqX $  & 10 & 5.7 & 2.3 
\\
\hline
CDF\&D{\O} &$\bar{p}p \rightarrow tb X$ & 2 & 1.9 & 1.1  
\\
& $\bar{p}p \rightarrow tqX $  &  &  &  
\\
\hline
ATLAS, CMS & $pp \rightarrow t\bar{t}X $ & 
  23 & 14 & 13
\\
\hline
CDF\&D{\O} &$\bar{p}p \rightarrow t\bar{t}X $   & 1 & 1.4 & 0.2
\\
\hline
\multicolumn{1}{l}{\bf Total} & \multicolumn{1}{c}{} & \multicolumn{1}{|c|}{2855} &\multicolumn{1}{c|}{3427} & \multicolumn{1}{c|}{3378}  \\ 
\cline{3-5}
  \end{tabular}
\caption{\small 
\label{tab:chi2}
The values of $\chi^2$ obtained in the present analysis at NLO for the data on 
inclusive DIS, the DY process, and on heavy-quark production in comparison with the 
ones of the ABMP16 fit at NNLO~\cite{Alekhin:2017kpj}.
}
\end{center}
\end{table}

%% file: table-cut.tex
\begin{table}[h!]
\renewcommand{\arraystretch}{1.3}
\fontsize{10.25}{11.25}\selectfont
\begin{center}
\begin{tabular}{|c|c|c|c|}   
\hline                           
\multicolumn{2}{|c|}{fit ansatz}                      
&\multicolumn{2}{c|}{$\alpha_s(M_Z)$}  
\\
\hline                                                    
$\quad$ {higher twist modeling} $\quad$
&{cuts on DIS data}                      
&$\quad$ {NLO} $\quad\quad\quad\quad$ 
&$\quad$ {NLO} $\quad\quad\quad$ 
\\                                                        
\hline                                                    
{higher twist fitted}
&$Q^2>2.5~{\rm GeV}^2$, $W>1.8~{\rm GeV}$ $\quad$
&0.1191(11)
&0.1147(8)\phantom{0}
\\\hline
\multirow{2}{*}
&$Q^2>10~{\rm GeV}^2$, $W^2>12.5~{\rm GeV}^2$ 
&0.1212(9)\phantom{0}
&0.1153(8)\phantom{0}
\\
\cline{2-4}
{higher twist fixed at 0} &$Q^2>15~{\rm GeV}^2$, $W^2>12.5~{\rm GeV}^2$
 &0.1201(11)
 &0.1141(10)
\\
\cline{2-4}
 &$Q^2>25~{\rm GeV}^2$, $W^2>12.5~{\rm GeV}^2$
 &0.1208(13)
 &0.1138(11)
\\
\hline                                          
\end{tabular}
\caption{\small 
  \label{tab:cuts}
  The values of $\alpha_s(M_Z)$ obtained in the NLO and NNLO variants 
  of the ABMP16 fit with various kinematic cuts on the DIS data 
  imposed and different modeling of the higher twist terms.}
\end{center}
\end{table}

%% file: table-mhq.tex
\begin{table}[h!]
\renewcommand{\arraystretch}{1.3}
\fontsize{10.25}{11.25}\selectfont
\begin{center}
\begin{tabular}{|l|l|l|}   
\hline                                                    
&$\quad$ {NLO} $\quad\quad\quad$ 
&$\quad$ {NNLO} $\quad\quad$ 
\\                                                        
\hline                                                    
 $\quad m_c(m_c)$~[GeV]$ \quad$
&$1.175\pm0.033$
&$1.252\pm0.018$
\\
\hline
 $\quad m_b(m_b)$~[GeV]$ \quad$
&\phantom{0}$3.88\pm0.13$ 
&\phantom{0}$3.84\pm0.12$
\\
\hline
 $\quad m_t(m_t)$~[GeV]$ \quad$
 &$162.1\pm1.0$
 &$160.9\pm1.1$
\\
\hline                                          
\end{tabular}
\caption{\small 
  \label{tab:hqmass}
The values of the $c$-, $b$- and $t$-quark masses in the \msbar\ scheme 
in units of GeV obtained in the NLO and NNLO variants of the ABMP16 fit. 
The quoted errors reflect the uncertainties in the analyzed data.   
}
\end{center}
\end{table}

%% file: table-asmz-mq-parameters-nlo.tex
\begin{table}[ht!]
\begin{center}
\renewcommand{\arraystretch}{1.3}
\begin{tabular}{|l|c|c|c|c|c|c|}
\hline
PDF set
  & $\alpha_s^{(n_f=5)}(M_Z)$
  & $m_c(m_c)$~[GeV]
  & $m_b(m_b)$~[GeV]
  & $m_b^{\rm pole}$~[GeV]
  & $m_t(m_t)$~[GeV]
  & $m_t^{\rm pole}$~[GeV]
\\[0.5ex]
\hline
  {\bf \phantom{0}0} &    {\bf 0.11905} &  {\bf 1.175} &   {\bf 3.880} &  {\bf 4.488} &   {\bf 162.08} &  {\bf 171.44}
\\[0.5ex]
\hline
\phantom{0}1 &  0.11905 & 1.175 & 3.880 & 4.488 & 162.08 & 171.44
 \\
\phantom{0}2 &  0.11906 & 1.175 & 3.880 & 4.489 & 162.08 & 171.44
 \\
\phantom{0}3 &  0.11905 & 1.175 & 3.880 & 4.488 & 162.08 & 171.44
 \\
\phantom{0}4 &  0.11899 & 1.175 & 3.880 & 4.488 & 162.08 & 171.43
 \\
\phantom{0}5 &  0.11898 & 1.175 & 3.880 & 4.487 & 162.08 & 171.43 
 \\
\phantom{0}6 &  0.11907 & 1.175 & 3.880 & 4.489 & 162.08 & 171.44
 \\
\phantom{0}7 &  0.11906 & 1.175 & 3.880 & 4.489 & 162.08 & 171.44
 \\
\phantom{0}8 &  0.11911 & 1.175 & 3.880 & 4.489 & 162.08 & 171.44
 \\
\phantom{0}9 &  0.11858 & 1.175 & 3.880 & 4.482 & 162.08 & 171.40
 \\
10 &  0.11925 & 1.175 & 3.880 & 4.491 & 162.08 & 171.46
 \\
11 &  0.11914 & 1.175 & 3.880 & 4.490 & 162.08 & 171.45
 \\
12 &  0.11910 & 1.176 & 3.880 & 4.489 & 162.08 & 171.44
 \\
13 &  0.11904 & 1.173 & 3.880 & 4.488 & 162.08 & 171.44
 \\
14 &  0.11909 & 1.203 & 3.880 & 4.489 & 162.08 & 171.44
 \\
15 &  0.11912 & 1.170 & 3.880 & 4.489 & 162.08 & 171.44
 \\
16 &  0.11930 & 1.169 & 3.877 & 4.489 & 162.08 & 171.46
 \\
17 &  0.11897 & 1.174 & 3.754 & 4.351 & 162.08 & 171.43
 \\
18 &  0.11883 & 1.179 & 3.878 & 4.483 & 162.09 & 171.43
 \\
19 &  0.11904 & 1.175 & 3.884 & 4.493 & 162.08 & 171.44
 \\
20 &  0.11879 & 1.180 & 3.888 & 4.493 & 162.13 & 171.47
 \\
21 &  0.11901 & 1.179 & 3.872 & 4.479 & 162.06 & 171.41
 \\
22 &  0.11914 & 1.180 & 3.882 & 4.492 & 162.05 & 171.41
 \\
23 &  0.11889 & 1.169 & 3.880 & 4.486 & 162.12 & 171.47
 \\
24 &  0.11879 & 1.178 & 3.875 & 4.479 & 161.86 & 171.19
 \\
25 &  0.11980 & 1.169 & 3.881 & 4.500 & 162.97 & 171.44
 \\
26 &  0.11914 & 1.182 & 3.881 & 4.491 & 162.12 & 171.49
 \\
27 &  0.11892 & 1.171 & 3.879 & 4.486 & 161.89 & 171.23
 \\
28 &  0.11888 & 1.176 & 3.882 & 4.488 & 161.88 & 171.21
 \\
29 &  0.11936 & 1.176 & 3.870 & 4.482 & 162.34 & 171.74
\\[0.5ex]
\hline
\end{tabular}
\caption{\small 
  \label{tab:parameters}
  Values of the heavy-quark masses 
  $m_c(m_c)$, $m_b(m_b)$ and $m_t(m_t)$ and 
  $\alpha_s^{(n_f=5)}(M_Z)$ 
  in the \msbar scheme for the PDFs {\tt ABMP16\_5\_nlo (0+29)} with $n_f=5$.    
  The values for pole masses $m_b^{\rm pole}$ and
  $m_t^{\rm pole}$ in the on-shell scheme obtained using {\tt RunDec}~\cite{Chetyrkin:2000yt} 
  are also given.
}
\end{center}
\end{table}

%% file: table-crs.tex
\begin{table}[t!]
\begin{center}
\renewcommand{\arraystretch}{1.3}
\begin{tabular}{|l||c||c|c|c|c|}
\hline
  & \multirow{2}{6em}{\, $\sigma(H)$~[pb] at \\ \, $\sqrt{s}=13$~TeV}
  & \multirow{2}{6em}{\, $\sigma(t{\bar t})$~[pb] at \\ \, $\sqrt{s}=5$~TeV}
  & \multirow{2}{6em}{\, $\sigma(t{\bar t})$~[pb] at \\ \, $\sqrt{s}=7$~TeV}
  & \multirow{2}{6em}{\, $\sigma(t{\bar t})$~[pb] at \\ \, $\sqrt{s}=8$~TeV}
  & \multirow{2}{6em}{\, $\sigma(t{\bar t})$~[pb] at \\ \, $\sqrt{s}=13$~TeV}
\\[3.5ex]
\hline
\phantom{N}NLO, {\tt ABMP16\_5\_nlo}
  & $ 33.59 \pm 0.58 $
  & $ 63.74  \pm  1.44 $
  & $ 172.0  \pm  3.3 $
  & $ 247.9  \pm  4.5 $
  & $ 835.3  \pm  14.3 $
\\[0.5ex]
\hline
NNLO, {\tt ABMP16\_5\_nnlo} 
  & $ 40.20 \pm 0.63 $
  & $ 63.66 \pm 1.60 $  
  & $ 171.8 \pm 3.4 $  
  & $ 247.5 \pm 4.6 $  
  & $ 831.4 \pm 14.5 $  
\\[0.5ex]
\hline
\end{tabular}
  \caption{\small 
  \label{tab:crs}
  Cross sections at NLO and NNLO in QCD
  for the Higgs boson production from gluon-gluon fusion 
  ($\sigma(H)$ computed in the effective theory) at $\sqrt{s}=13$~TeV for $m_H=125.0$~GeV
  with the renormalization and factorization scales set to $\mu_r=\mu_f=m_H$ 
  and for the top-quark pair production, $\sigma(t{\bar t})$, 
  at various center-of-mass energies of the LHC with 
  the top-quark mass $m_t(m_t)$ in the \msbar scheme and $\mu_r=\mu_f=m_t(m_t)$. 
  The values of $\alpha_s(M_Z)$ and $m_t(m_t)$ are order dependent, 
  see Eq.~(\ref{eq:as}) and Tab.~\ref{tab:hqmass}. 
  The errors denote the PDF and $\alpha_s$ uncertainties.
  }
\end{center}
\end{table}

%% file: nlo-journalversion.bbl
\begin{thebibliography}{10}

\bibitem{Accardi:2016ndt}
A.~Accardi {\em et~al.},
\newblock Eur. Phys. J. {\bf C76}, 471 (2016), arXiv:1603.08906.

\bibitem{Alekhin:2012ig}
S.~Alekhin, J.~Bl{\"u}mlein, and S.~Moch,
\newblock Phys. Rev. {\bf D86}, 054009 (2012), arXiv:1202.2281.

\bibitem{Alekhin:2013nda}
S.~Alekhin, J.~Bl{\"u}mlein, and S.~Moch,
\newblock Phys. Rev. {\bf D89}, 054028 (2014), arXiv:1310.3059.

\bibitem{Alekhin:2017kpj}
S.~Alekhin, J.~Bl{\"u}mlein, S.~Moch, and R.~Placakyte,
\newblock Phys. Rev. {\bf D96}, 014011 (2017), 
\hfill \\
arXiv:1701.05838.

\bibitem{Buckley:2014ana}
A.~Buckley {\em et~al.},
\newblock Eur. Phys. J. {\bf C75}, 132 (2015), arXiv:1412.7420.

\bibitem{Blumlein:1996gv}
J.~Bl{\"u}mlein, S.~Riemersma, W.~L. van Neerven, and A.~Vogt,
\newblock Nucl. Phys. Proc. Suppl. {\bf 51C}, 97 (1996), arXiv:hep-ph/9609217.

\bibitem{Alekhin:2016evh}
  S.~Alekhin, J.~Bl{\"u}mlein and S.~O.~Moch,
  Mod.\ Phys.\ Lett.\ A {\bf 31} (2016) no.25,  1630023.

\bibitem{Andreev:2017vxu}
  H1 Collaboration, V.~Andreev {\em et al.},
  Eur.\ Phys.\ J.\ C {\bf 77} (2017) no.11,  791, arXiv:1709.07251.

\bibitem{Rojo:2015acz}
J.~Rojo {\em et~al.},
\newblock J. Phys. {\bf G42}, 103103 (2015), arXiv:1507.00556.

\bibitem{Dulat:2015mca}
S.~Dulat {\em et~al.},
\newblock Phys. Rev. {\bf D93}, 033006 (2016), arXiv:1506.07443.

\bibitem{Ball:2017nwa}
NNPDF, R.~D. Ball {\em et~al.},
\newblock Eur. Phys. J. {\bf C77}, 663 (2017), arXiv:1706.00428.

\bibitem{Prytz:1993vr}
  K.~Prytz,
\newblock  Phys.\ Lett.\ B {\bf 311}, 286 (1993).

\bibitem{Harland-Lang:2015qea}
  L.~A.~Harland-Lang, A.~D.~Martin, P.~Motylinski and R.~S.~Thorne,
  Eur.\ Phys.\ J.\ C {\bf 76} (2016) no.1,  10, arXiv:1510.02332.

\bibitem{Marquard:2015qpa}
  P.~Marquard, A.~V.~Smirnov, V.~A.~Smirnov and M.~Steinhauser,
  Phys.\ Rev.\ Lett.\  {\bf 114},  142002 (2015),
  arXiv:1502.01030.

\bibitem{Harlander:2002wh}
R.~V. Harlander and W.~B. Kilgore,
\newblock Phys. Rev. Lett. {\bf 88}, 201801 (2002), arXiv:hep-ph/0201206.

\bibitem{Anastasiou:2002yz}
C.~Anastasiou and K.~Melnikov,
\newblock Nucl. Phys. {\bf B646}, 220 (2002), arXiv:hep-ph/0207004.

\bibitem{Ravindran:2003um}
V.~Ravindran, J.~Smith, and W.~L. van Neerven,
\newblock Nucl. Phys. {\bf B665}, 325 (2003), arXiv:hep-ph/0302135.

\bibitem{Aliev:2010zk}
M.~Aliev {\em et~al.},
\newblock Comput. Phys. Commun. {\bf 182}, 1034 (2011), arXiv:1007.1327.

\bibitem{Baernreuther:2012ws}
P.~B{\"a}rnreuther, M.~Czakon, and A.~Mitov,
\newblock Phys. Rev. Lett. {\bf 109}, 132001 (2012), arXiv:1204.5201.

\bibitem{Czakon:2012zr}
M.~Czakon and A.~Mitov,
\newblock JHEP {\bf 12}, 054 (2012), arXiv:1207.0236.

\bibitem{Czakon:2012pz}
M.~Czakon and A.~Mitov,
\newblock JHEP {\bf 01}, 080 (2013), arXiv:1210.6832.

\bibitem{Czakon:2013goa}
M.~Czakon, P.~Fiedler, and A.~Mitov,
\newblock Phys. Rev. Lett. {\bf 110}, 252004 (2013), arXiv:1303.6254.

\bibitem{Chetyrkin:2000yt}
K.~G. Chetyrkin, J.~H. K{\"u}hn, and M.~Steinhauser,
\newblock Comput. Phys. Commun. {\bf 133}, 43 (2000), arXiv:hep-ph/0004189.

\end{thebibliography}
